\documentclass[pra,showpacs,superscriptaddress,10pt,twocolumn]{revtex4-2}
\usepackage{amsmath,amssymb,graphicx}
\usepackage{epstopdf}
\usepackage{xcolor}
\begin{document}

\title{Nonlocal Activation of Bound Entanglement via Local Quantum Zeno Dynamics}

\author{Fatih Ozaydin}\email{fatih@tiu.ac.jp}
\affiliation{Institute for International Strategy, Tokyo International University, 1-13-1 Matoba-kita, Kawagoe, Saitama 350-1197, Japan}

\author{Cihan Bayindir}
\email{cbayindir@itu.edu.tr}
\affiliation{Engineering Faculty, Istanbul Technical University,  34469 Maslak, Istanbul, Turkey}
\affiliation{Engineering Faculty, Bogazici University, 34342 Bebek, Istanbul, Turkey}
\affiliation{CERN, 1211 Geneva 23, Switzerland}

\author{Azmi Ali Altintas}
\email{altintas.azmiali@gmail.com}
\affiliation{Department of Physics, Faculty of Science, Istanbul University, 34116 Vezneciler, Istanbul, Turkey}

\author{Can Yesilyurt}
\email{can\texttt{-{}-}yesilyurt@hotmail.com}
\affiliation{Institute of Nanotechnology and Biotechnology, Istanbul University-Cerrahpasa, 34320 Istanbul, Turkey }

\date{\today}

\begin{abstract}
Bound entanglement was shown to be activated [P. Horodecki \textit{et al.,} Phys. Rev. Lett. \textbf{82,} 1056 (1999)] in the sense that the entanglement of a spatially separated two-qutrit system can be increased with nonzero probability via a sufficiently large number of preshared bound-entangled states, local three-level controlled operations, and classical communications. 
Here, we present a local quantum Zeno scheme for activating bound entanglement which is based only on single-particle rotations and threshold measurements. 
In our scheme, neither a large number of bound-entangled states nor controlled operations are required, and classical communication is required only once at the end of the protocol. 
We show that a single bound-entangled state is sufficient for increasing the negativity of the target entangled state from 0.11 to 0.17, and by using four more bound-entangled states, negativity can be made greater than 0.42 and the fidelity to the maximally entangled state increases from 0.3 to 0.41, 0.50, 0.59, and 0.61.
We believe our results are important not only for quantum technologies but also for a better understanding of quantum entanglement.
\end{abstract}


\maketitle

\section{Introduction}

	A quantum system $\rho$ consisting of two subsystems $\rho^A$ and $\rho^B$ is inseparable if it cannot be written in the form
	\begin{equation}
		\rho = \sum_i p_i \rho_i^A \otimes \rho_i^B
	\end{equation}
	
	\noindent with $\sum_i p_i = 1$
and a fundamental aspect of inseparable quantum states is that they can exist in distillable 
or nondistillable forms. 
That is, out of an arbitrary number of inseparable states, the question is whether or not finite entanglement can be distilled~\cite{NC,Horodecki1997PLA,Horodecki1998PRLIsThere,Horodecki2001QIP,Bennett1999PRL} with only stochastic local operations and classical communications.

The answer is negative for inseparable states with positive partial transpose (PPT), i.e. bound entangled (BE) states.
The emergence of a BE state can be explained through an irreversible process in the sense that an initial entanglement is required for it to emerge, but once emerged, no entanglement can be distilled from it.
While qubit-qubit and qubit-qutrit systems satisfy the separability criterion of Peres stating that states with PPT are separable, this is not the case for higher dimensions~\cite{Horodecki1998PRLIsThere}. 
Hence, being in the smallest dimension that bound entanglement can be observed, qutrit-qutrit systems deserve a particular attention for understanding the fundamentals of quantum mechanics and quantum information.

\indent
Besides the fact that no entanglement can be distilled from them, it was shown that BE states are not \textit{useless} in the sense that they can enable tasks which are impossible to realize with separable states, or provide advantages over them.
For example, a secret key can be distilled from BE states~\cite{Horodecki2005PRL,Horodecki2008IEEE}.
BE states can be useful for quantum metrology~\cite{Horodecki2015PRA}, and also in quantum thermodynamics~\cite{Horodecki02PRL,Tuncer2019QIP}.
Hence, BE states are important not only for quantum science and technology, but also for understanding quantum entanglement.\\
\indent Another interesting feature of BE states is that they can be activated~\cite{Horodecki1999PRL}.
That is, let Alice and Bob initially share a sufficiently large number of BE states and one non-maximally free entangled (FE) state. 
Taking a BE state in each iteration, they apply local controlled-operations between the particles of the BE state and the FE state that they share. 
Measuring the particles of the BE state and communicating the results over a classical channel, entanglement of the FE state is increased by some probability.
They discard that BE state, and taking another, continue the iteration until ending up with a maximally  entangled state.
The drawback of this scenario is that it requires a sufficiently large number of BE states (which requires initial entanglement), controlled gates, and classical communications.
Furthermore, this scenario is based on qutrits and realizing three-level controlled gates is less practical and more error prone than two-level controlled gates, due to physical imperfections, which decreases the overall success probability and the fidelity of the target state.

\onecolumngrid
\onecolumngrid
\begin{figure}[t!]
	\centerline{\includegraphics[width=1\columnwidth]{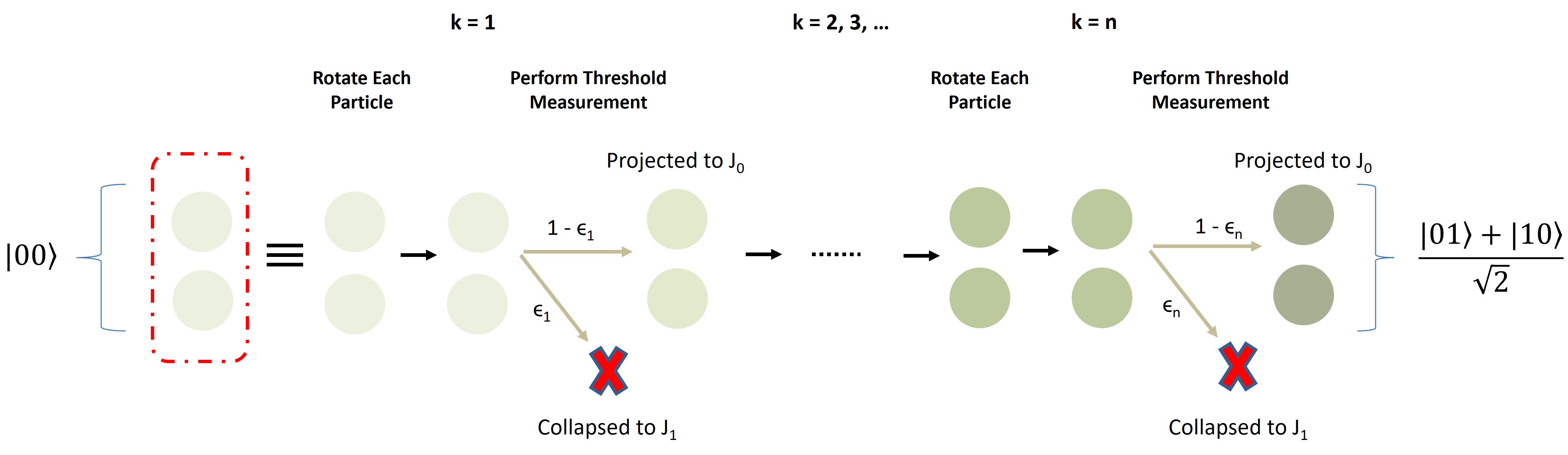}}
	\caption{Steps of the entanglement creation strategy of Ref.~\cite{Nori08PRA}.
			In each iteration, after rotating each qubit with a small angle, a threshold measurement (the so-called $J-$measurement) is performed with two projectors $J_1 = |1\rangle\langle1|\otimes|1\rangle\langle1|$ and $J_0 = \mathcal{I} - J_1$. With a small probability $\epsilon_k$, the state collapses to $J_1$ and the protocol fails. Otherwise, with probability $1-\epsilon_k$, the state is projected to the $J_0$ subspace. After $n$ iterations, initially separable particles in the $|00\rangle$ state can be found in the maximally entangled state $(|01\rangle + |10\rangle) /\sqrt{2}$, as if a controlled-NOT operation was implemented. }\label{fig:Nori_strategy_steps}
\end{figure}
\twocolumngrid

However, realizing controlled gates is not the only way to manipulate multiparticle quantum systems.
Wang \textit{et al.} showed that it is possible to create entanglement between two qubits via a quantum Zeno effect consisting of single particle rotations and frequent threshold measurements~\cite{Nori08PRA}, which would require controlled gates or Bell measurements otherwise~\cite{NC}.
The quantum Zeno effect was first considered for slowing down and even freezing the evolution of a quantum system by inhibiting a certain quantum subspace through properly designed frequent measurements~\cite{SudarshanMisra1977}.

Consider a quantum system in state $|1\rangle$ at $t=0$. 
The probability to find the system in the same state later at $t>0$ is
\begin{equation}
	p(t) = \left| \left\langle e \left|\exp \left( -{i\over \hbar} \hat{H} t\right) \right| e\right\rangle \right|^2,
\end{equation}	

\noindent where $\hat{H}$ is the governing Hamiltonian.
For a Hamiltonian with finite variance $\langle V^2 \rangle$ and short times, the probability is then $p(t) \approx 1- \langle V^2 \rangle t^2 / \hbar^2 $.
Performing projective measurements at intervals $\tau$ with a sufficiently large frequency $1/\tau \gg \langle V^2 \rangle ^{1 \over 2} / \hbar$, we find 
$p^n(\tau) = p(t = n t) \approx \exp [ - ( \langle V^2 \rangle \tau / \hbar^2) t ]$.
Hence, the decay of the system from the initial state $|1\rangle$ slows down with $\tau$, and with $\tau \rightarrow 0$, the evolution of the system is frozen, which is known as the quantum Zeno effect.

Itano \textit{et al.} experimentally observed the collapse of wave function back to its initial state~\cite{Itano90PRA}.
Since then, because the major problem in quantum technologies is the decay of the quantum systems due to inevitable interactions with the environment, in addition to error correcting codes~\cite{Cory98PRL,Chiaverini04Nature} intense theoretical and experimental efforts have been devoted to slow down the evolution of the system via the quantum Zeno effect in various settings from cavity quantum electrodynamics~\cite{Bernu08PRL,Raimond10PRL,Raimond12PRA} to large atomic systems~\cite{Signoles14NatPhys}, cold atoms~\cite{Fischer01PRL}, ion traps~\cite{Beige04PRA}, nuclear magnetic resonance systems~\cite{Zheng13PRA}, and Bose-Einstein condensates~\cite{Schafer14NatComm}.
We have recently showed that frequent measurements can freeze rogue waves~\cite{Bayindir18OptComm}, and similar to inhibiting a quantum subspace, the motion of quantum chirps can be inhibited in certain domains~\cite{Bayindir21PLA}.
Kofman and Kurizki showed that in addition to slowing it down, appropriately designed frequent measurements can also accelerate the quantum decay of a system~\cite{Kurizki00Nature}, which could provide advantages in various quantum tasks, such as in realizing quantum heat engines~\cite{Kurizki20CommPhys}. 
Because preparing multi-partite entanglement in specific forms such as $W$ states requires sophisticated methods and realization of several controlled gates~\cite{OurW13PRA,OurW13QIP,OurW14PRA,OurW16SciRep,OurW16JOSAB,OurW15OptExp,OurW20SciRep,OurW21PRA}, quantum Zeno schemes were proposed~\cite{Chen16OptComm} and experimentally realized~\cite{Barontini17Science} to solve this problem.

On the other hand, the Dzyaloshinskii–Moriya interaction~\cite{Dzyaloshinskii,Moriya}, which is shown to excite not only entanglement~\cite{ZhangDM07PRA} but also quantum Fisher information~\cite{OurDM15,OurDM20} and to be a fast quantum information scrambler~\cite{OurDMScramble} was recently proposed for freeing bound entanglement~\cite{Sharma16QIP}. 
Besides several application areas in quantum information and computation, consideration of the quantum Zeno effect for activating bound entanglement is lacking.

As illustrated in Fig.~\ref{fig:Nori_strategy_steps}, Wang et al. showed that a maximally entangled state can be prepared out of two initially separable qubits without implementing a controlled operation or a Bell measurement~\cite{Nori08PRA}. 
Entanglement is rather created by a protocol consisting of a repeated sequence of \textit{rotate-measure} actions. 
That is, starting from the initial $|0\rangle|0\rangle$ state, in each iteration, following single qubit rotations with a small angle, a two-qubit collective measurement with projectors $J_1 = |1\rangle\langle1|\otimes|1\rangle\langle1|$ and $J_0 = \mathcal{I} - J_1$ is performed. 
The state collapses to $J_1$ subspace and the protocol fails with a small probability.
Otherwise, the state is projected to $J_0$ subspace, i.e. the evolution of the system to the $|11\rangle$ state is \textit{inhibited} via frequent measurements.
After a number of iterations, initially separable particles in the state $|00\rangle$ can be projected to the maximally entangled state $(|01\rangle + |10\rangle) /\sqrt{2}$, almost \textit{deterministically}. 
In other words, realizing the evolution $|00\rangle \rightarrow (|01\rangle + |10\rangle) /\sqrt{2}$, quantum Zeno dynamics consisting of single-particle rotations and frequent threshold measurements can be considered as an alternative to implementing a controlled-operation between two qubits.

In this paper, considering the difficulty in realizing a large number of three-level controlled gates in the original activation proposal~\cite{Horodecki1999PRL} and motivated by the work of Wang \textit{et al.}~\cite{Nori08PRA}, 
we ask whether it is possible with a finite probability to activate bound entanglement via quantum Zeno dynamics,  i.e., to connect two interesting phenomena of nature for an operational purpose.
Hence, the contribution of the present paper is twofold. 
First, we show that it is possible to activate bound entanglement via local quantum Zeno dynamics. Second, we show that, unlike the original protocol, the activation can be realized without implementing three-level controlled quantum operations.

A basic difference from the present scheme is that in Ref.~\cite{Nori08PRA}, entanglement was created between a pair of initially separable qubits via quantum Zeno dynamics.
However, in our scheme, as illustrated in Fig.~\ref{fig:strategy} and detailed in Fig.~\ref{fig:strategy_steps}, we consider a bound-entangled state and a free but nonmaximally entangled state shared by spatially separated parties. 
Each party applies the local quantum Zeno dynamics on the particles of the shared bound- and free-entangled states.
A carefully designed quantum Zeno strategy can be an alternative to implementing local controlled-XOR operations on the two qutrits held by each party in the work of Horodecki \textit{et al.} for activating bound entangled, i.e. using the BE states, the entanglement of the FE state is excited~\cite{Horodecki1999PRL}.

We consider the bipartite bound-entangled state described in the three-dimensional Hilbert space spanned by basis states $\{|0\rangle=(1,  0,  0 )^T, |1\rangle=(0,  1,  0 )^T, |2\rangle=(0,  0,  1 )^T \} $
as in Ref.~\cite{Horodecki1999PRL},

\begin{equation}\label{eq:sigma_alpha}
	\sigma_{\alpha} = {2 \over 7 } |\Psi_+ \rangle \langle \Psi_+| + {\alpha \over 7 } \sigma_{+} + {5-\alpha \over 7 } \sigma_{-},
 \end{equation}

\noindent where

\begin{equation}\label{eq:PsiP} 
	|\Psi_+ \rangle = { |00\rangle +  |11\rangle + |22\rangle \over \sqrt{3} },  
\end{equation}

\begin{equation} \sigma_+ = {  |01 \rangle \langle 01| + |12 \rangle \langle 12| + |20 \rangle \langle 20| \over 3 },  \end{equation}

and
\begin{equation} \sigma_- = {  |10 \rangle \langle 10| + |21 \rangle \langle 21| + |02 \rangle \langle 02| \over 3 }.  \end{equation}

This state is separable for $2 \leq \alpha \leq 3$, bound entangled for $3 < \alpha \leq 4$ and free entangled for $4 < \alpha \leq 5$.
In the original superactivation scenario ~\cite{Horodecki1999PRL}, Alice and Bob take the initially shared free but non-maximally entangled state 
\begin{equation}\label{eq:sigma_free}
	\sigma_{\text{free}} = F  |\Psi_{+} \rangle \langle \Psi_{+} |  + (1-F) \sigma_{+} 
\end{equation} 

\noindent with $0<F<1$, and one of the BE pairs $\sigma_{\alpha}$.
Each applies the unitary $\text{XOR}$ operation $U_{\text{XOR}} |a\rangle |b\rangle = |a\rangle |b \oplus a \rangle$, with $b\otimes a = (b \otimes a) \text{mod} N$ ~\cite{Bennett96PRL}, the $N$-level particle of the free (bound) pair being the source (target).
Each measures the particle of the target (BE) pair on the $z$-axis and they communicate the result. 
If the result differs, they discard the target pair without affecting the source pair. 
Otherwise, the fidelity of the resulting free-entangled state $\rho_{out}$ with the maximally entangled state 
$F(\rho_{out}) = \langle \Psi_{+}| \rho_{\text{out}} |\Psi_{+} \rangle$
is improved as 
$F'(F) = 2 F / [2F + (1-F)(5-\alpha)]$ 
with probability 
$P_{F \rightarrow F'} = {1 \over 7} [2F + (1-F)(5-\alpha)]$.\\

\section{Activating Bound Entanglement via Quantum Zeno Dynamics}
\begin{figure}[t!]
	\centerline{\includegraphics[width=0.85\columnwidth]{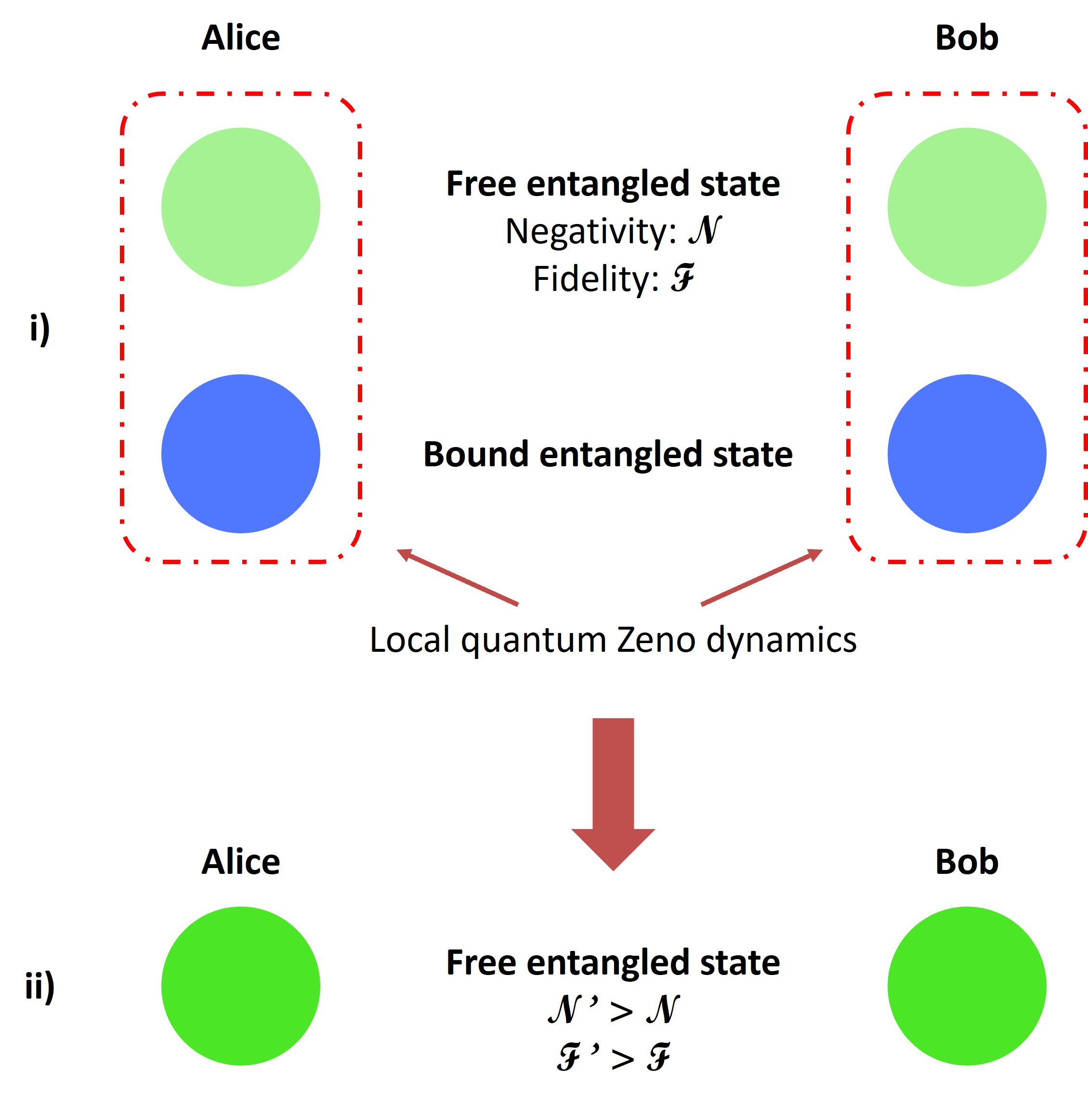}}
	\caption{Strategy for activating bound entanglement via quantum Zeno dynamics. 
		i) Alice and Bob initially share a nonmaximally free-entangled state
		with  negativity $N$ and a bound-entangled state.  
		Each applies a local quantum Zeno strategy 
		consisting of only single-particle rotations and threshold measurements, measure the particle of the initially bound-entangled state, and communicates the result. 
		ii) With a nonzero probability, they are left with the free-entangled state 
		with negativity $N' > N$ and  $F' > F$. See the text and Fig.~\ref{fig:strategy_steps} for details. }\label{fig:strategy}
\end{figure}

\begin{figure}[t!]
	\centerline{\includegraphics[width=0.9\columnwidth]{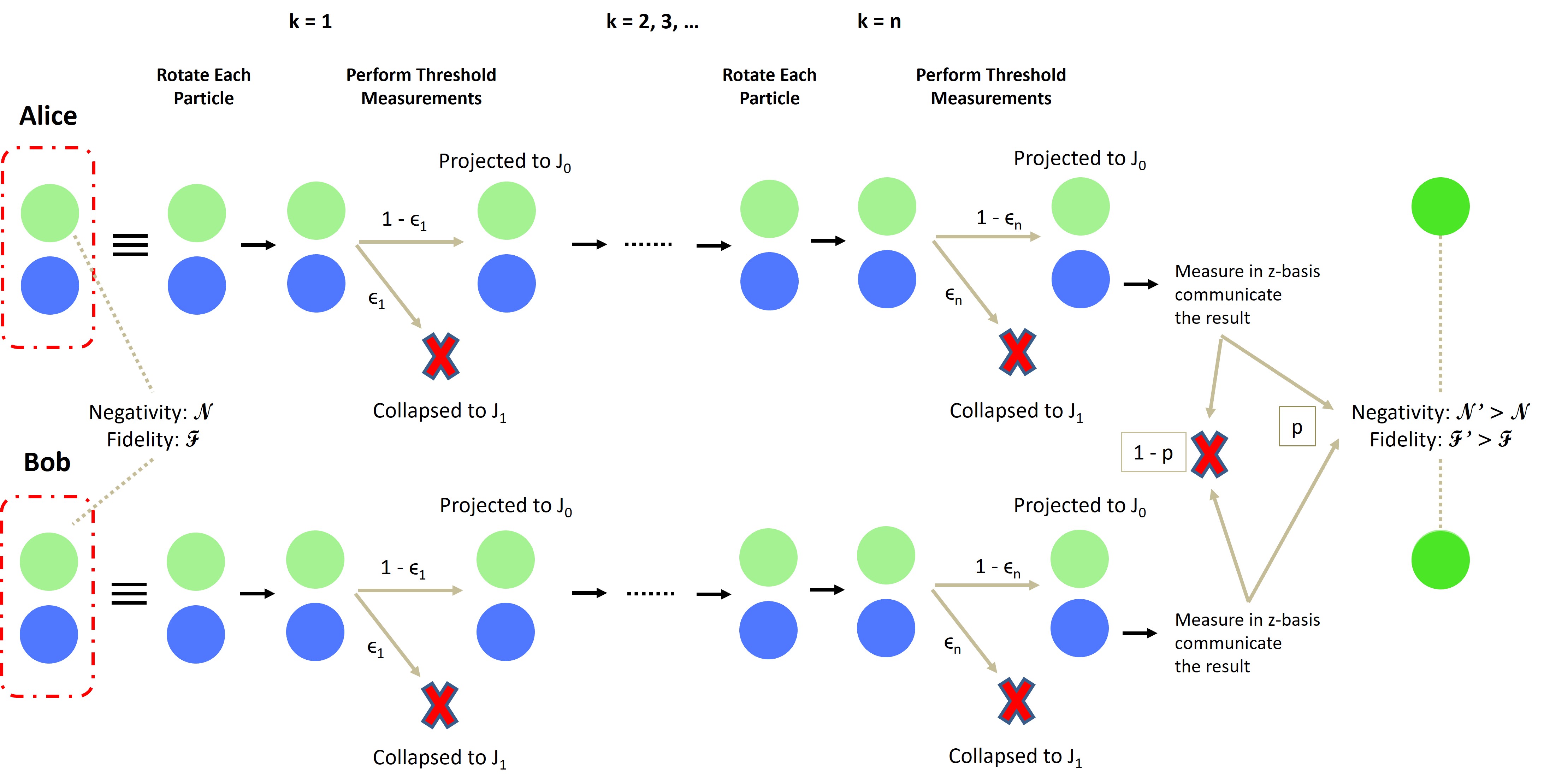}}
	\caption{Steps of our activation strategy illustrated in Fig.~\ref{fig:strategy} detailed. 
			In each iteration, after rotating each particle with a small angle, threshold measurements are performed. With a small probability $\epsilon_k$, the states collapse to $J_1$ and the protocol fails. With probability $1-\epsilon_k$, states are projected to $J_0$ subspace. After $n$ iterations, particles of the initially BE state are measured on $z$ basis and the results are communicated. With probability $p$, results differ and the protocol fail; with probability $1-p$, the protocol successfully improves the initially FE state's negativity and the fidelity to the maximally entangled state.}\label{fig:strategy_steps}
\end{figure}

In our scheme as illustrated in Fig.~\ref{fig:strategy}, each party repeats the \textit{rotate-measure} actions as the main operation for sufficiently many times (to be determined by numerical simulations) and then they measure the particles of the initial BE state, leaving the FE state with a higher entanglement and an improved fidelity with a finite probability.

Applying single-particle rotations on each particle by a small angle around the $x$, $y$, or $z$ axis or their combinations, say, $R(\theta)$ for simplicity, i.e.,

\begin{equation}
\rho = R^{\otimes 4}(\theta) (\sigma_{\text{free}} \otimes \sigma_{\alpha} ) (R^{\dagger})^{\otimes 4}(\theta),
\end{equation}

\noindent each party performs the threshold measurement on their two particle systems with two projectors

\begin{equation} \label{eq:Js}
J_1= |i\rangle\langle i| \otimes |j\rangle\langle j|, \ \ \ \ \ J_0 = \mathcal{I}-J_1,
\end{equation}

\noindent where $i, j \in \{0,1,2\} $ and $\mathcal{I}$ is the nine-dimensional identity operator.

As explained in Ref.~\cite{Nori08PRA} with an exemplary implementation by a Josephson-junction circuit with flux qubits (see Fig.1 of Ref.~\cite{Nori08PRA}), this threshold detection, or the so-called $J-$measurement is not a measurement in the $z-$ basis finding each particle in some state, but rather, it detects whether or not the particles are in the $|i\rangle|j\rangle$ state.
Hence, if the two particles are found in the $|i\rangle|j\rangle$ state with a small probability $\epsilon$, the system collapses to the corresponding $J_1$ subspace. 
Otherwise, the system is projected to the $J_0 = \mathcal{I}-J_1$ subspace. 
Qiu \textit{et al.} showed in Ref.~\cite{Qiu16SRep} that, due to significant separation of the third energy level from the higher levels, a four-junction circuit can also be used as a quantum three-level system (qutrit). 
Therefore, the threshold measurements can be implemented in four-junction superconducting circuits, similar to the threshold measurement in Ref.~\cite{Nori08PRA}.

Projecting the system to the $J_0 = \mathcal{I}-J_1$ subspace, its new state is

\begin{equation} \label{eq:new_state}
\rho' = {(J_0 \otimes J_0) \ \rho \ (J_0 \otimes J_0)^{\dagger} \over \text{Tr}[(J_0 \otimes J_0) \ \rho \ (J_0 \otimes J_0)^{\dagger}]}.
\end{equation}

\noindent After repeating the above \textit{rotate-measure} operation $n$ times, each party measures the particle of the initially bound-entangled state in hand and they are left with the $\sigma'_{\text{free}}$ state. 
Here the parameters $n$, $\theta$, $i$, and $j$ and the final measurement scenario on the particles of the initially BE state are to be determined.

Below we will show via the results of our numerical simulation that the negativity of $\sigma'_{\text{free}}$ and its fidelity to the maximally entangled state $|\Psi^+\rangle$
can be increased, i.e., $N(\sigma'_{\text{free}}) > N(\sigma_{\text{free}})$ and $F(\sigma'_{\text{free}}) > F(\sigma_{\text{free}})$, with a simple strategy.
In our simulation, we choose a single rotation axis for simplicity, i.e., rotation around the $z$ axis 

\begin{equation}
\displaystyle R(\theta) = 
\left(
\begin{array}{ccc}
\cos \theta & - \sin \theta & 0 \\
\sin \theta & \ \ \cos \theta & 0 \\
0 & 0 & 1 \\
\end{array}
\right),
\label{eq:HRZ}
\end{equation}

\noindent and we set the small rotation angle to be $\theta=\pi/180$ on each particle in each iteration. 
We also assume the same $\{J_1 , J_0\}$ measurements for both parties and for each iteration. 
Hence, our simulation's main goal is to determine the set of parameters $\{i,j\}$ for the measurement operators, the number of iterations $n$, and the measurement result on the particles of an initially BE state which lead to an increase in the negativity of the FE state.
Even in this simple setting via fixing those parameters for each iteration, we are able find a strategy for activating the bound entanglement with a finite probability.
The negativity of a bipartite state $\rho$ is calculated via the absolute sum of its negative eigenvalues $\mu_i$ of its partial transpose $\rho^{\Gamma_A}$ with respect to subsystem $A$, 
	or equivalently as $N(\rho) \equiv { ||\rho^{\Gamma_A}||_1 - 1 \over 2}$ where $||X||_1$ is the trace norm of the operator $X$~\cite{Eisert99JOM,Erol14SRep,Vidal2002PRA}. We also calculate the fidelity of the obtained state $\sigma$ to the maximally entangled state $|\Psi_{+}\rangle$ given in Eq.(2), i.e., $F(\sigma) = \langle \Psi_{+}| \sigma | \Psi_{+} \rangle$ as in the original superactivation scenario~\cite{Horodecki1999PRL}.
We start with the free entangled state with $F=0.3$, $N(\sigma_{\text{free}})\approxeq 0.11$, and $\alpha=4$ for $\sigma_{\alpha}$, which is a BE state. 
For each possible $J_1= |i\rangle\langle i| \otimes |j\rangle\langle j|$ with $i, j \in \{0,1,2\} $ for both parties in each iteration, we perform the \textit{rotate-measure} action. 
We run the simulation $n$ times and at the $k$th run with $k=1,2,...,n$, particles which initially belonged to the BE state are measured on the $z$ axis. 
For each of nine possible results, we calculate the negativity of the resulting FE state $\sigma'_{\text{free}}$ to check whether it is greater than the negativity of the initial FE state $\sigma_{\text{free}}$.

As shown in Fig. \ref{fig:result0}, through threshold measurements $\{J_1 , J_0\}$ followed by the best final $z$-basis measurement outcomes, we find that after a number of iterations between 242 and 282 with the peak value at 262, and $J_1= |0\rangle\langle 0| \otimes |1\rangle\langle 1|$ for both parties and finding both particles of the initially BE state in state $|1\rangle$, the negativity of the state written in the computational basis

\begin{equation}
	\displaystyle \sigma'_{\text{free}} \!=\!\! 
	\left(
	\begin{array}{ccccccccc}
		0 & 0 & 0 & 0 &	0 & 0 & 0 & 0  &	0 \\
		0 & 0 &  0 & 0 &	0 & 0 & 0 & 0  &	0 \\
		0 & 0 &  0 & 0 &	0 & 0 & 0 & 0  &	0 \\
		0 & 0 &  0 & 0 &	0 & 0 & 0 & 0  &	0 \\
		0 & 0 & 0 & 0 &	0.29649 & 0 & 0 & 0  &	0.30824 \\
		0 & 0 &  0 & 0 &	0 & 0.38043 & 0 & 0  &	0 \\
		0 & 0 &  0 & 0 &	0 & 0 & 0 & 0   &	0 \\
		0 & 0 &  0 & 0 &	0 & 0 & 0 & 0.00104  &	0 \\
		0 & 0 & 0 & 0 &	0.30824 & 0 & 0 & 0  & 0.32202 \\
	\end{array}
	\right)
	\label{eq:sigma'_free}
\end{equation}

\noindent is made $0.171 195$. 
Hence, increasing the negativity from approximately $0.11$ to approximately $0.17$, we show that bound entanglement can be activated via quantum Zeno dynamics. 
We find that the probability of projecting the target state to the desired subspace is greater than $0.78$ in the first iteration and greater than $0.999$ in all the remaining iterations, clearly indicating the quantum Zeno dynamics. 
For a better understanding of the exact evolution of the system, in the Appendix we provide an explicit calculation for one iteration.
The drawback of our scheme with this simple parameter set is that, following the quantum Zeno dynamics, the probability of finding the particles of the initially BE state in state  $|1\rangle$ is $0.04$, which we discuss how to improve in the next section. 

\onecolumngrid
\onecolumngrid
\twocolumngrid

Through numerical simulations, we also find that repeating exactly the same procedure with a fresh BE state each time and the obtained free-entangled state from the previous run, 
the states with density matrices

	\begin{equation}
		\displaystyle \sigma''_{\text{free}} \!=\!\! 
		\left(
		\begin{array}{ccccccccc}
			0 & 0 & 0 & 0 &	0 & 0 & 0 & 0  &	0 \\
			0 & 0 &  0 & 0 &	0 & 0 & 0 & 0  &	0 \\
			0 & 0 &  0 & 0 &	0 & 0 & 0 & 0  &	0 \\
			0 & 0 &  0 & 0 &	0 & 0 & 0 & 0  &	0 \\
			0 & 0 & 0 & 0 &	0.34417 & 0 & 0 & 0  &	0.37414 \\
			0 & 0 &  0 & 0 &	0 & 0.244553 & 0 & 0  &	0 \\
			0 & 0 &  0 & 0 &	0 & 0 & 0 & 0   &	0 \\
			0 & 0 &  0 & 0 &	0 & 0 & 0 & 0.00242  &	0 \\
			0 & 0 & 0 & 0 &	0.37414 & 0 & 0 & 0  & 0.40884 \\
		\end{array}
		\right)
		\label{eq:sigma''_free}
	\end{equation}
	
	\begin{equation}
		\displaystyle \sigma'''_{\text{free}} \!=\!\! 
		\left(
		\begin{array}{ccccccccc}
			0 & 0 & 0 & 0 &	0 & 0 & 0 & 0  &	0 \\
			0 & 0 &  0 & 0 &	0 & 0 & 0 & 0  &	0 \\
			0 & 0 &  0 & 0 &	0 & 0 & 0 & 0  &	0 \\
			0 & 0 &  0 & 0 &	0 & 0 & 0 & 0  &	0 \\
			0 & 0 & 0 & 0 &	0.37490 & 0 & 0 & 0  &	0.44561 \\
			0 & 0 &  0 & 0 &	0 & 0.08169 & 0 & 0  &	0 \\
			0 & 0 &  0 & 0 &	0 & 0 & 0 & 0   &	0 \\
			0 & 0 &  0 & 0 &	0 & 0 & 0 & 0.01062  &	0 \\
			0 & 0 & 0 & 0 &	0.44561 & 0 & 0 & 0  & 0.53277 \\
		\end{array}
		\right)
		\label{eq:sigma'''_free}
	\end{equation}
	
	\noindent and 
	
	\begin{equation}
		\displaystyle \sigma''''_{\text{free}} \!=\!\! 
		\left(
		\begin{array}{ccccccccc}
			0 & 0 & 0 & 0 &	0 & 0 & 0 & 0  &	0 \\
			0 & 0 &  0 & 0 &	0 & 0 & 0 & 0  &	0 \\
			0 & 0 &  0 & 0 &	0 & 0 & 0 & 0  &	0 \\
			0 & 0 &  0 & 0 &	0 & 0 & 0 & 0  &	0 \\
			0 & 0 & 0 & 0 &	0.36606 & 0 & 0 & 0  &	0.45496 \\
			0 & 0 &  0 & 0 &	0 & 0.04417 & 0 & 0  &	0 \\
			0 & 0 &  0 & 0 &	0 & 0 & 0 & 0   &	0 \\
			0 & 0 &  0 & 0 &	0 & 0 & 0 & 0.02078  &	0 \\
			0 & 0 & 0 & 0 &	0.45496 & 0 & 0 & 0  & 0.56897 \\
		\end{array}
		\right)
		\label{eq:sigma''''_free}
	\end{equation}

\noindent are obtained in each round with negativity and fidelities presented in Table~\ref{T1}.

To create entanglement from two separable particles such as in Ref.~\cite{Nori08PRA}, an arbitrary input state can be selected, and the Zeno dynamics can be designed accordingly.
However in the present work, the input states are determined by the parameters $F$ and $\alpha$.
Following the original activation work by Horodecki \textit{et al.}~\cite{Horodecki1999PRL}, we choose $F = 0.3$ to have a comparative perspective on two proposals. 
For the $\sigma_{\alpha}$ state, our simulation results are in agreement with Ref.~\cite{Horodecki1999PRL}. 
That is, in the separable region $2 \leq \alpha \leq 3$, no activation is possible via quantum Zeno dynamics.
In the region $3 < \alpha \leq 5$, the greater $\alpha$ is, the better the performance achieved (not necessarily with the same number of iterations $n$). 
However, because our focus is on the bound entangled region $3 < \alpha \leq 4$, we choose $\alpha = 4$ to achieve the best performance.
The choice of small rotation angle $\theta$ does not significantly change the performance of the protocol in this simple design, and the trivial choice $\theta = \pi / 180$ achieves a decent performance. 
However, as we will discuss in the next section, considering possibly different $\theta$ values for each iteration by each party in a quantum evolutionary algorithm might improve the performance.
Regarding the number of iterations, any value in a wide range is sufficient to provide a proof of concept for activation via local quantum Zeno dynamics. 
However, we find that 262 iterations achieve the best performance in the current setting.

\begin{table}[t!]
	\centering
		\caption{Negativity and fidelity to the maximally entangled $|\Psi_+\rangle$ of the initial state $\sigma_{\text{free}}$ and the states obtained after each round of local Zeno dynamics (Fig.~\ref{fig:strategy}) with a fresh BE state.}\label{T1}
	\begin{tabular}{c c c c}
		\hline
		\hline	
		\ \ \ \ \ Round \ \ \ \ \	&  \ \ \ State \ \ \ 	&   \ \ \ Negativity \ \ \ & \ \ \ Fidelity \\ [1ex]				
		\hline \\
			&	$\sigma_{\text{free}}$		& 0.110977		& 0.3  	  \\ [1ex]
		1	&  	$\sigma'_{\text{free}}$		& 0.171195		& 0.411667			  \\ [1ex]
		2	&  	$\sigma''_{\text{free}}$	& 0.269747		& 0.500432	  \\ [1ex]
		3 	&  	$\sigma'''_{\text{free}}$	& 0.400867		& 0.599635 	  \\ [1ex]
		4	&  	$\sigma''''_{\text{free}}$	& 0.422634		& 0.614989	  \\ [1ex]
		\hline
		\hline
	\end{tabular}

\end{table}

\section{Discussion}

\begin{figure}[t!]
	\centerline{\includegraphics[width=1\columnwidth]{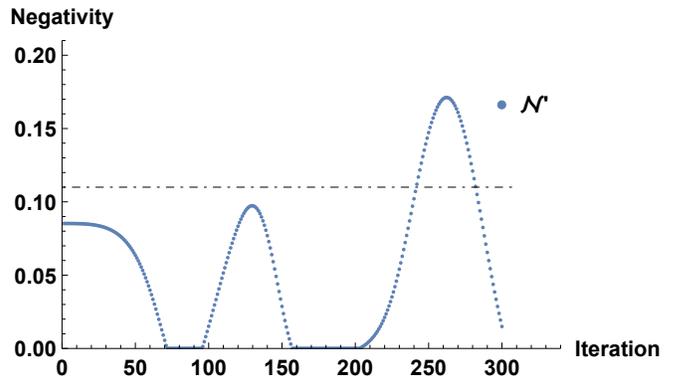}}
	\caption{Activating bound entanglement via local quantum Zeno dynamics. The dashed-dotted line at approximately $0.11$ shows $N$, the negativity of the initial FE state.	Using a single BE state and performing the rotate-measure operation $k$ times, each point of the curve shows $N'$, the negativity of the final FE state, which achieves the maximum value approximately equal to $0.17$ for $k=262$, upon measuring the particles of the initially BE state in $z$-axis and finding both in the $|1\rangle$ state.}\label{fig:result0}
\end{figure}

As can be seen in Fig.~\ref{fig:result0}, the negativity of the resulting state is not fixed at a point, i.e., the state is not \textit{frozen} via quantum Zeno dynamics. 
What is frozen, or \textit{inhibited} is a specific subspace, leaving the complementary subspace \textit{uninhibited.}
Within the uninhibited subspace, the state continues to evolve, and the evolution can be controlled by quantum Zeno dynamics.
Hence, fluctuations and even periodic behavior in some features of quantum systems can be observed during a continuous quantum Zeno dynamics.

In Fig.~\ref{fig:result0} the curve for $N'$ does not start at the negativity value of the initial state, i.e., approximately equal to $0.11$ because a rotation is already applied on each particle and the result shows the case where the particle of the initial BE state is found in a specific state. 
The negativity value hits zero in two bands as $k\rightarrow100$ and $k\rightarrow200$ implying that the initial FE state turns to be either a separable or a bound-entangled state and then back to a FE state.
As both cases are interesting and it is not straightforward to check whether an arbitrary state is a separable or a BE state, we note this problem as an open question for future research.

On the one hand, acting on the same input states, both the present and the original protocol~\cite{Horodecki1999PRL} lead to a similar result in activating the BE state by increasing the negativity of the FE state and its fidelity to the maximally entangled state.
On the other hand, the present quantum Zeno dynamics is not directly implementing an operation similar to a three-level XOR operation.
Furthermore, the original protocol increases the fidelity by preserving the structure of the density matrix bringing it closer to the maximally entangled state $|\Psi^+\rangle = (|00\rangle + |11\rangle + |22\rangle)/\sqrt{3}$, whereas the present protocol does not, so a reliable comparison between the present operation and the XOR operation does not seem to be applicable. 
Hence, considering the similar case in Ref.~\cite{Nori08PRA}, i.e. realizing the evolution $|00\rangle \rightarrow (|01\rangle + |10\rangle)/\sqrt{2}$ without applying a controlled-NOT gate, one can consider the existence of different routes in quantum protocols.

With the same initial states and parameters, i.e. $F= 0.3$ and $\alpha = 0.4$, the original activation protocol~\cite{Horodecki1999PRL} can increase the fidelity to the maximally entangled state from $0.3$ to $0.46$ and then to $0.63$, while the present protocol can increase it to $0.41$, and then to $0.50$. 
Also considering the implementation of a single XOR operation versus a series of rotate-measure actions in each round, the improvement via the original protocol is faster than the present one.
However, the significant advantage of the present protocol is that the physical realization of the operations it requires is more practical.
The present protocol requires single-particle rotations and simple threshold measurements which are easy to implement with high fidelity, while the original protocol requires two three-level controlled operations in each round, and it is still a challenge in various technologies to implement a three-level quantum operation with high fidelity.

Because our quantum Zeno strategy can be regarded as an alternative to implementing the XOR operations considered in Ref.~\cite{Horodecki1999PRL}
theoretically, there is no limit on increasing the entanglement of the FE state to the maximum value, and
our results suggest that with a better design, the initially non-maximally entangled state can be made close to a maximally entangled state via quantum Zeno dynamics.

Furthermore, in our simple design, bound entanglement is activated if the measurement outcomes of both BE particles are $|1\rangle$ only, which obviously leads to the small success probability $0.04$. 
Hence with a better parameter set and possibly a more sophisticated strategy such as an intelligent evolution of~\cite{Nori08PRA}, it might be expected that in a decent case, not a single but more possible outcomes can lead to activation with a greater total success probability. 
In the best case, any outcome could result in activation. 
Hence, rather than measuring the particles on the $z$ axis and communicating the results via a classical channel, Alice and Bob could simply discard the particles of the BE state after the final rotate-measure operation. 
However in a general scenario, there are three rotation axes and a large set of possible values for small rotation angles for each of the four particles, as well as 81 possible $J$ measurements for each iteration and finally nine possible outcomes when measuring the BE particles on the $z$ axis. 
This implies that in order to find an optimal strategy via a brute-force approach, almost $18 \times 10^6$ possibilities need to be analyzed for around 300 iterations.
Furthermore, the number of iterations to reach the optimal result is not known \textit{a priori} and in the present work it is determined by numerical simulations.
Therefore, as future research for finding an optimal set of parameters, designing a quantum evolutionary algorithm appears to be a good candidate where individuals represent the instances of these parameters and the fitness function is based on the entanglement measure.
Also, it would be interesting to develop a machine learning algorithm to predict the number of iterations in such scenarios.


\section{Conclusion}
We have proposed a local quantum Zeno scheme for non-local activation of bound entanglement of a spatially separated system.
The advantages over the previous activation proposal~\cite{Horodecki1999PRL} are that (i) our scheme does not require any three-level controlled-operations on the particles, (ii) classical communication is required only once and at the end of the frequent rotate-measure operations, and (iii) a single BE state has the potential to increase the negativity of the FE state from $0.11$ to $0.17$, and by using four more BE states, the negativity of the target FE state can be made greater than $0.27$, $0.35$, $0.40$ and $0.42$, respectively, 
and the fidelity of the FE state to the maximally entangled state can increase from $0.3$ to $0.41$, $0.50$, $0.59$, and $0.61$.\\
\indent The drawback of our proposed scheme is that via the simple parameter set, the final success probability is small. 
In order to find an optimal parameter set and a better strategy, due to the huge search space of the parameters, advanced artificial intelligence techniques might be required.
Hence, as a proof of concept work, we presented a simple scheme and discussed a potential evolutionary quantum algorithm to improve it.\\
\indent Because bound entanglement is one of the most interesting and counter-intuitive phenomena of quantum mechanics and so is the Zeno effect, we believe that bringing these two together and using the quantum Zeno effect to activate bound entanglement have the potential not only to enable advances in quantum technologies, but also to contribute to a better understanding of quantum entanglement.

\section*{Acknowledgments}
F.O. acknowledges financial support from Tokyo International University Personal Research Fund. C.Y. acknowledges support from the Istanbul University Scientific Research Fund through Grant No. BAP-2019-33825.\\

\onecolumngrid
\onecolumngrid
\section*{Appendix}
\setcounter{equation}{0}
\renewcommand{\theequation}{A.\arabic{equation}}

We will now explicitly show the evolution of the free-entangled state in the first iteration.
For $F=0.3$, the initial free-entangled state defined in Eq.(\ref{eq:sigma_free}) is

\begin{equation} \label{eq:sigma_f_0.3}
	\displaystyle \sigma_{\text{free}} \!=\!\! 
	\left(
	\begin{array}{ccccccccc}
		0.1 & 0 	& 0 & 0 &	0.1 & 0 	& 0 	& 0  &	0.1 \\
		0   & 0.233 & 0 & 0 &	0   & 0 	& 0 	& 0  &	0 \\
		0   & 0 	& 0 & 0 &	0   & 0 	& 0 	& 0  &	0 \\
		0   & 0 	& 0 & 0 &	0   & 0 	& 0 	& 0  &	0 \\
		0.1 & 0 	& 0 & 0 &	0.1 & 0 	& 0 	& 0  &	0.1 \\
		0   & 0 	& 0 & 0 &	0   & 0.233 & 0 	& 0  &	0 \\
		0   & 0 	& 0 & 0 &	0   & 0 	& 0.233 & 0  &	0 \\
		0   & 0 	& 0 & 0 &	0   & 0 	& 0 	& 0  &	0 \\
		0.1 & 0 	& 0 & 0 &	0.1 & 0 	& 0 	& 0  &  0.1 \\
	\end{array}
	\right),	
\end{equation}

\noindent
yielding fidelity $F(\sigma_{\text{free}})=0.3$ to the maximally entangled state $|\Psi_+\rangle$ defined in Eq.(\ref{eq:PsiP}).
For $\alpha=4$, the initial bound-entangled state defined in Eq.(\ref{eq:sigma_alpha}) is

\begin{equation} \label{eq:sigma_alfa_4}
	\displaystyle \sigma_{\alpha} \!=\!\! 
	\left(
	\begin{array}{ccccccccc}
		0.095238 	& 0 		& 0 		& 0 		&	0.095238 	& 0 		& 0  		& 0 		& 0.095238 \\
		0   	  	& 0.190476 	& 0 		& 0 		&	0   		& 0 		& 0 		& 0  		& 0 		\\
		0   	  	& 0 		& 0.047619 	& 0 		&	0  			& 0 		& 0 		& 0  		& 0 		\\
		0   	  	& 0 		& 0 		& 0.047619 	&	0   		& 0 		& 0 		& 0  		& 0 		\\
		0.095238 	& 0 		& 0 		& 0 		&	0.095238 	& 0 		& 0  		& 0 		& 0.095238 	\\
		0   	  	& 0 		& 0 		& 0 		&	0   		& 0.190476 	& 0 		& 0  		&	0 		\\
		0   		& 0 		& 0 		& 0 		&	0   		& 0 		& 0.190476 	& 0  		&	0 		\\
		0   		& 0 		& 0 		& 0 		&	0   		& 0 		& 0 		& 0.047619 	&	0 		\\
		0.095238 	& 0 		& 0 		& 0 		&	0.095238 	& 0 		& 0 		& 0  		&  0.095238	\\
	\end{array}
	\right).	
\end{equation}

The first iteration starts with applying the single particle rotation defined in Eq.(\ref{eq:HRZ}) by a small angle $\theta=\pi/180$, i.e.,

\begin{equation}
	\displaystyle R(\pi/180) = 
	\left(
	\begin{array}{ccc}
		0.999848 	& -0.0174524 	& 0 \\
		0.0174524 	& \ \ 0.999848 	& 0 \\
		0 			& 0 			& 1 \\
	\end{array}
	\right),
	\label{eq:HRZ_value}
\end{equation}

\noindent
on each particle. 
Next, each party performs the threshold measurement with the projectors $J_1$ and $J_0$ given in Eq. (\ref{eq:Js}) on the possessed two-particle system, one being the particle of the FE and the other being the a particle of the BE system, hoping that the systems are projected to $J_0$ subspace.

If the protocol is stopped right after the first iteration and each party performs the final measurement on the particle of the BE state and communicates the result, with a nonzero probability, the free-entangled system will be in the state

\begin{equation} \label{eq:rho_after_first_iteration}
	\displaystyle \sigma^1_{\text{free}} \!=\!\! 
	\left(
	\begin{array}{ccccccccc}
		0 			& 0 		& 0 		& 0 		&	0 			& 0 		& 0  		& 0 		& 0 \\
		0   	  	& 0 		& 0 		& 0 		&	0   		& 0 		& 0 		& 0  		& 0 		\\
		0   	  	& 0 		& 0 		& 0 		&	0  			& 0 		& 0 		& 0  		& 0 		\\
		0   	  	& 0 		& 0 		& 0 		&	0   		& 0 		& 0 		& 0  		& 0 		\\
		0 			& 0 		& 0 		& 0 		&	0.230895 	& 0 		& 0  		& 0 		& 0.230731 	\\
		0   	  	& 0 		& 0 		& 0 		&	0   		& 0.538209 	& 0 		& 0  		&	0 		\\
		0   		& 0 		& 0 		& 0 		&	0   		& 0 		& 0 		& 0  		&	0 		\\
		0   		& 0 		& 0 		& 0 		&	0   		& 0 		& 0 		& 0.047619 	&	0 		\\
		0 			& 0 		& 0 		& 0 		&	0.230731 	& 0 		& 0 		& 0  		&  0.230731	\\
	\end{array}
	\right).	
\end{equation}

\noindent
The fidelity of the $\sigma^1_{\text{free}}$ state to the maximally entangled state $|\Psi_+\rangle$ is $F(\sigma^1_{\text{free}})=0.307 696$, which shows an improvement over the fidelity of the initial $\sigma_{\text{free}}$ state $F(\sigma_{\text{free}})=0.3$.
If the protocol runs for 262 iterations before the final measurement and classical communication, the $\sigma^{262}_{\text{free}} = \sigma'_\text{free}$ state in Eq.(\ref{eq:sigma'_free}) would be obtained, achieving fidelity $F(\sigma^{262}_{\text{free}})=0.411 667$.

\twocolumngrid

\end{document}